\documentclass[submission,copyright,creativecommons]{eptcs}

\providecommand{\event}{TFPIE 2022} 
\usepackage[english]{babel}
\usepackage[utf8]{inputenc}
\usepackage[final]{microtype}
\usepackage{multicol}
\usepackage{xcolor}
\usepackage{graphicx}
\usepackage{float}
\usepackage{subcaption}
\usepackage{pgfplots}

\usepackage{url}
\definecolor{pastelblue}{RGB}{0,72,205}
\hypersetup{
 colorlinks,
 linktoc=page,
 linkcolor=pastelblue,
 citecolor=pastelblue,
 urlcolor=pastelblue
}
\usepackage[inline]{enumitem}
\usepackage[outputdir=build,cachedir=minted-cache,frozencache,cachedir=minted-cache]{minted}
\BeforeBeginEnvironment{minted}{\smallskip}
\AfterEndEnvironment{minted}{\smallskip}

\usepackage{listings}

\lstdefinelanguage{cyp}{
  keywords={Lemma,goal,Proof,QED,To,show,Case},
  keywordstyle=\color{blue}\bfseries,
  keywords=[2]{data},
  keywordstyle=[2]{\color{purple}\bfseries},
  keywords=[3]{.=.},
  keywordstyle=[3]{\color{red}\bfseries},
  comment=[l]{--},
  commentstyle=\color{purple}\ttfamily,
}

\lstdefinestyle{cyp}{
  language=cyp,
  extendedchars=true,
  basicstyle=\ttfamily,
  showstringspaces=false,
  showspaces=false,
  tabsize=2,
  breaklines=true,
  showtabs=false
}

\lstdefinestyle{Haskell}{
  language=Haskell,
  extendedchars=true,
  basicstyle=\ttfamily,
  showstringspaces=false,
  showspaces=false,
  tabsize=2,
  breaklines=true,
  showtabs=false
}

\usepackage{tikz}
\usetikzlibrary{quotes,angles,calc,arrows.meta,positioning,automata}
\tikzset{initial text={}}
\usepackage{tikz-3dplot}

\usepackage[capitalize,sort,compress,noabbrev]{cleveref}
\crefdefaultlabelformat{#2\textup{#1}#3}
\Crefname{thm}{Theorem}{Theorems}
\Crefname{lem}{Lemma}{Lemmas}
\Crefname{prop}{Proposition}{Propositions}
\Crefname{cor}{Corollary}{Corollary}
\Crefname{defn}{Definition}{Definitions}
\Crefname{exmpl}{Example}{Examples}
\Crefname{rmk}{Remark}{Remarks}
\Crefname{propy}{Property}{Properties}
\Crefname{claim}{Claim}{Claim}
\Crefname{assmlisti}{Assumption}{Assumptions}
\Crefname{invarlisti}{Invariant}{Invariants}
\usepackage{titlesec}
\titlespacing*{\paragraph}{0pt}{0.5\baselineskip plus \baselineskip minus 0.2\baselineskip}{\baselineskip}

\title{Engaging, Large-Scale Functional Programming Education in Physical and Virtual Space}

\author{Kevin Kappelmann \qquad\qquad Jonas Rädle \qquad\qquad Lukas Stevens
\institute{Department of Informatics\\ Technical University of Munich, Germany}
\email{\quad kevin.kappelmann@tum.de \quad\qquad raedle@in.tum.de \quad\qquad stevensl@in.tum.de}
}
\def\titlerunning{Engaging, Large-Scale Functional Programming Education}
\def\authorrunning{K. Kappelmann, J. Rädle \& L. Stevens}

\begin{document}
\maketitle

\begin{abstract}
  Worldwide, computer science departments have experienced a dramatic increase in the number of student enrolments.
Moreover, the ongoing COVID-19 pandemic requires institutions to radically replace the traditional way of on-site teaching,
moving interaction from physical to virtual space.
We report on our strategies and experience tackling these issues
as part of a Haskell-based functional programming and verification course,
accommodating over 2000 students in the course of two semesters.
Among other things,
we fostered engagement with weekly programming competitions
and creative homework projects,
workshops with industry partners,
and collaborative pair-programming tutorials.
To offer such an extensive programme to hundreds of students,
we automated feedback for programming as well as
inductive proof exercises.
We explain and share our tools and exercises so that they can be reused by other educators.

\end{abstract}


\section{Introduction}\label{sec:intro}

This paper reports on strategies and tools employed to
run two iterations of a large-scale functional programming and verification course at the Technical University of Munich (TUM).
While the first iteration (winter semester 2019, 1057 participants)
took place on campus,
the second iteration (winter semester 2020, 1031 participants) was affected by the COVID-19 pandemic and took place in virtual space.
Previous iterations of the course were introduced in~\cite{next_1100};
however, we were facing two novel challenges:

\paragraph{Soaring Enrolments}
The relatively young field of computer science has
become one of the largest study programmes around the globe.
The increase of student enrolments is dramatic
\cite{comp_sci_growth1,comp_sci_growth2}
while employment of new teaching staff often lags behind.
At TUM, the number of new enrolments in computer science more than doubled between 2013 and 2021 from 1110 to 2644 (an increase of 138\%)
while academic staff only increased from 439 to 573 (31\%) \cite{tum_numbers}.

This drastic increase not only requires more physical resources -- like larger lecture halls and more library spaces --
but also academic staff for supervision.
Given the discrepancy in growth between student enrolments and staff employment,
automation of supervision and feedback mechanisms is inevitable.
Automation, however, should not
negatively affect the quality of the teaching.

\paragraph{Online Teaching}
The ongoing COVID-19 pandemic forced a radical
transition from on-site teaching to online classes.
Lecturers had to rethink the way they present material and interact with students,
teaching assistants the way they assist students in tutorial sessions.
Students, on the other hand, suffer from a lack of social interaction and communication, leading to higher
levels of stress, anxiety, loneliness, symptoms of depression, and diminished affective engagement~\cite{students_lockdown1,onlineengagement1}.

In our experience, the disconnect between students and lecturers as well as the lack of on campus interaction between students may also lead to \emph{cramming}:
the practice of showing little participation during the semester
while studying extensively just before the exam.
Cramming tends to result in poor long-term retention and shallow understanding of material.
Indeed, the benefit of spacing learning events apart rather than cramming has been demonstrated in hundreds of experiments \cite{cramming1,cramming2}.

\vspace{0.5\baselineskip}\noindent
Besides these general challenges,
there is a third -- subject-specific --
challenge we were keen to tackle:

\paragraph{Functional Programming is Practical}
Feedback by students and personal experience has shown us that many students
at TUM question the applicability and usefulness
of functional languages beyond academia.
They are disappointed by a lack of industrial insight
and real world -- or at least interactive -- applications.
Indeed, some even perceive functional programming as an obstacle;
after all, they already know how to program imperatively.

Good educators do not just teach but inspire:
we have to bring the benefits of functional languages
closer to our students' hearts
by showing real-world applicability and making functional programming fun and engaging.

\paragraph{Contributions}

In this paper,
we present our answers to these challenges
and provide tools and exercises for other educators.
Our contributions are
\begin{enumerate*}[label=\arabic*)]
  \item a description of technical tools that enable large-scale (functional) programming courses in physical and virtual space,
  \item a toolbox of engagement mechanisms
for programming courses
and an evaluation thereof, and
  \item reusable exercises and tools for other functional programming educators.
\end{enumerate*}
Our resources can be found in the central repository of this paper.\footnote{\url{https://github.com/kappelmann/engaging-large-scale-functional-programming/}}

\paragraph{Outline}

We begin by describing the underlying conditions of our course and its syllabus in \cref{sec:course_structure_conditions}.
The tools and teaching methods employed during lectures are explained in \cref{sec:lectures}.
\cref{sec:practical_part} describes the
mechanisms, tools, and technical setup
we used to create an engaging experience
that is scalable
for the practical part of the course.
It also includes a framework for I/O
testing in Haskell
that solves a transparency issue prevalent in traditional approaches.
\cref{sec:cyp} introduces
``Check Your Proof'' -- a tool created
by our lab to automatically check simple inductive proofs for Haskell programs.
\cref{sec:exam} describes how we adapted our exams to the COVID-19 situation and large number of participants.
Finally, \cref{sec:related_work} outlines related work
and \cref{sec:conclusion} concludes with a summary and aspects to improve.

\section{Course Structure and Conditions}\label{sec:course_structure_conditions}

\subsection{Conditions}

The 5~ECTS\footnote{European Credit Transfer System; one ECTS credit equals 30 hours of work} course was mandatory for computer science undergraduates in their third semester and
an elective for other related degrees such as games engineering or information systems.
All students studied Java in their first semester and had taken courses on algorithms and data structures,
discrete mathematics, and linear algebra.
The course ran for 14 weeks with
one 90-minute lecture,
one 90-minute tutorial,
and one exercise sheet each week.

1057 students registered for
the first iteration\footnote{\url{https://www21.in.tum.de/teaching/fpv/WS19/} (website -- except ``Wettbewerb'' -- German; course material English)} that took place on campus in winter semester 2019 (WS19) and
1031 for the second iteration\footnote{\url{https://www21.in.tum.de/teaching/fpv/WS20/} (English)} in winter semester 2020 (WS20), taking place in virtual space due to the COVID-19 pandemic.

Both iterations were organised by the lecturer, Tobias Nipkow, and the authors of this paper.
The former designed the lecture material, created the slides\footnote{\url{https://www21.in.tum.de/teaching/fpv/WS20/assets/slides.pdf}}, and delivered the lectures.
The others took care of the practical and organisational part of the course.
All gained valuable experience in running an online course on the theory of computation for 1071
students in summer semester 2020.
Finally, Manuel Eberl had the honour of assisting us with the weekly programming competition (see \cref{sec:engagement}).

Needless to say,
running tutorials for more than 1000
students each semester on our own is impossible.
We were further assisted by
13 student assistants in WS19 and
22 student assistants in WS20.
In WS19, their primary job was to run the tutorials and provide feedback for homework submissions (e.g.\ code quality).
However, it became clear to us
that this manual feedback is not effective
and that their time is better spent creating engaging exercises (see \cref{sec:engagement}).
On average, each assistant worked 10 hours per week,
spending 4 hours on running tutorials
and the remaining 6 hours
preparing themselves for the tutorials,
grading student submissions,
and implementing new exercises for students.

\subsection{Syllabus\label{sec:syllabus}}

The course deals with the basics of functional programming and the verification of functional programs.
Most parts of the course could be done using any functional language.
We chose Haskell because of its simple syntax, large user community, and good testing facilities (in particular QuickCheck).
The syllabus stayed close to the one presented in~\cite{next_1100}.
The changes are the omission of the parser case study, the rigorous introduction of computation induction and type inference, and the decision to split off I/O from monads and introduce it earlier.
The last is done in an effort to convince
students more quickly that pure functional languages can be practical and deal with side effects.
For ease of reference, we list the syllabus below.
New or modified topics are marked ($\ast$):

\begin{multicols}{2}
\begin{enumerate}
  \item Introduction to functional programming
  \item Basic Haskell: \mintinline{Haskell}{Bool}, QuickCheck, \mintinline{Haskell}{Integer} and \mintinline{Haskell}{Int}, guarded equations, recursion on numbers \mintinline{Haskell}{Char}, \mintinline{Haskell}{String}, tuples
  \item List comprehension, polymorphism, a glimpse of the Prelude, basic typeclasses (\mintinline{Haskell}{Num}, \mintinline{Haskell}{Eq}, \mintinline{Haskell}{Ord}), pattern matching, recursion on lists (including accumulating parameters), scoping by example
  \item Proof by structural induction and computation induction on lists ($\ast$)
  \item Type inference algorithm ($\ast$)
  \item Higher-order functions: \mintinline{Haskell}{map},~\mintinline{Haskell}{filter},~\mintinline{Haskell}{foldr}, $\lambda$-abstractions, extensionality, currying
  \item Typeclasses
  \item Algebraic datatypes and structural induction
  \item Concrete I/O without introducing monads ($\ast$)
  \item Modules: module syntax, data abstraction, correctness proofs
  \item Case studies: Huffman codings, skew heaps
  \item Lazy evaluation and infinite lists
  \item Complexity and optimisation
  \item Monads ($\ast$)
\end{enumerate}
\end{multicols}

Concepts are introduced in small, self-contained steps.
Characteristic features of functional programming languages such as
higher-order functions and algebraic data types are
only introduced midway through the course.
This makes the design of interesting practical tasks harder
but ensures that students are not overwhelmed by the diversity
of new principles that are not part of introductory imperative programming courses.
In general, the course progresses from ideas close to what is known from imperative languages (e.g.\ boolean conditions, recursion on numbers, auxiliary functions, etc.)
to simple applications of new concepts (e.g.\ recursion and induction on lists)
to generalised new concepts (e.g.\ algebraic data types and structural induction).

\section{Lectures}\label{sec:lectures}

We used a mix of slides,
live coding, and whiteboard proofs for the lectures.
Each topic came with small case studies and examples.
The latter were accompanied by suitable QuickCheck tests
and inductive correctness proofs when appropriate.
The proofs stayed close to the format accepted by the proof checker that was used in the practical part of the course (see~\cref{sec:cyp}).
The students particularly enjoyed the live coding sections,
which received 16 positive and no negative comments in the course evaluation form.
This is in line with prior studies,
which moreover confirm the effectiveness of live coding~\cite{livecoding1,livecoding2}.

In both iterations, students were allowed to interact and ask questions at any time.
However, synchronous interaction with hundreds of students is challenging:
While the lecturer cannot answer all questions due to time constraints,
many students are also too reserved to ask questions given the large audience.
As the semester progresses, interaction tends to degrade to
questions posed by a small community of motivated students;
questions shared by a majority, on the other hand, often go unheard.

In WS19, we partly addressed this problem by offering an asynchronous Q\&A forum\footnote{We used a Zulip instance hosted by our department: \url{https://zulip.com/case-studies/tum/}}
where students could post anonymously or using their real name.
The forum contained separate sections for the theoretical part (including the lectures)
and the practical part of the course.
Questions posted in the former were answered by the lecturer
to increase interaction between students and the lecturer -- the lack of which was often criticised by students in our department.
Questions in the latter were also answered by teaching assistants and other students.
Answers by students were explicitly encouraged by us and
outstanding contributions awarded a special prize at the end of the semester.

However, while the forum was a great success with more than 3800 posts per semester,
engagement in the theoretical section stayed far behind its practical counterpart ($\leq$ 2\%).
Moreover, it does not fully address the live interaction problem since
questions are answered asynchronously.
As such, students stuck with conceptional problems may not be able
to keep up with the rest of the lecture,
leading to frustration.

For the second iteration of the course -- the online semester --
we thus added a new interaction method.
The lectures were livestreamed and
interaction was made possible by means of a live Q\&A board\footnote{We used tweedback \url{https://tweedback.de/}}.
The board was moderated by a PhD student
sitting in the same room as the lecturer.
Questions could be answered and voted on by students as well as the moderator.
The moderator approved and answered very specific and simple questions directly while forwarding questions of general interest to the lecturer in order to increase engagement between the students and the lecturer.

We can report that this format increased engagement when compared to the first iteration:
students were less reluctant to submit questions because
\begin{enumerate*}[label=\arabic*)]
  \item they had the chance to ask questions anonymously,
  \item they were not afraid to ``interrupt'' the lecture, and
  \item they were able to ask new kinds of questions.
\end{enumerate*}
Examples of the third include discussion of alternative solutions by students,
organisational questions,
and slightly off-topic discussions that nevertheless increase engagement and curiosity.
On average, 20--30 students
submitted at least one message per lecture -- a significant improvement when compared to the number of interactions during typical offline lectures.
The effectiveness of synchronous interactions during lectures has been indicated in prior studies~\cite{onlineengagement3,onlineengagement1}.
We hence recommend to offer a moderated live Q\&A board even for lectures taking place on campus or running in a hybrid format.

Following the livestream,
the recordings were uploaded for asynchronous consumption.
Students watching the lectures asynchronously still had the chance to submit questions to the forum.

\section{Practical Part}\label{sec:practical_part}
\subsection{Engagement Mechanisms}\label{sec:engagement}

Students spend the majority of their time on the practical part of the course.
This is where they apply the theory explained in the lecture to tutorial and homework exercises in the form of programming tasks, proof exercises, and miscellaneous other assignments (type inference, transformation of programs into tail-recursive form, etc.).
As each student has unique interests, strengths and weaknesses, and different levels of commitment,
we employed a diverse set of mechanisms to engage them.

As outlined in the introduction, keeping up engagement is particularly challenging in courses that are taught remotely.
We experienced this first hand when teaching a course in theoretical computer science during the first semester affected by the COVID-19 pandemic.
We saw a significantly larger decrease in homework and tutorial participation over the course of that semester than in previous years.
We thus put a particular emphasis on engaging teaching methods for the functional programming course in WS20.

We want to emphasise that engagement does not simply
increase by offering more things -- this may even increase stress --
but by offering things that serve neglected needs.
Effective engagement mechanisms do not simply keep students busy
but genuinely make the course more interesting and fun:
they engage students with the content,
the instructors,
and with each other~\cite{onlineengagement3,onlineengagement2,engagementproposals}.
We now describe the mechanisms that proved particularly valuable to us:

\paragraph{Grade Bonus}
For both course iterations,
students were able to obtain a bonus of one grade step on their final exam provided that they achieved certain goals during the semester.
This incentive was already used in a previous iteration
but subsequently dropped due to negative experiences with plagiarism.
However, as a result, participation in homework exercises
severely decreased~\cite{next_1100}.
Moreover, the student council reported to us that one of the most asked for
wishes by students is that of a grade bonus.

We hence re-introduced the bonus with some changes.
First, instead of asking for $40\%$ of all achievable points,
we changed to a pass-or-fail per exercise sheet system.
Students passed a sheet if they
passed $\approx 70\%$ of all tests
and obtained the bonus if they passed $\approx 70\%$ of all sheets.
We changed to this system so that students
could not obtain the grade bonus early on in the semester and then stop participating,
leading to cramming.
Secondly, in WS20, we introduced additional ways to
obtain bonus points,
for example by participating in programming contests or workshops by industry partners.
This diversified the system and
particularly increased engagement of
students who were struggling
with programming tasks but were nevertheless interested in the course.

As a result, out of 802 students that interacted with the homework system,
298 obtained the grade bonus in WS20 ($37\%$).
More than $96\%$ of all students that obtained the bonus
passed the final exam,
whereas more than half of all students that did not obtain the bonus failed.
Similar numbers can be reported for WS19.
Finally, in contrast to previous years,
we have not seen any severe cases of plagiarism despite
running all submissions through a plagiarism checking tool\footnote{We used Moss \url{https://theory.stanford.edu/~aiken/moss/}}.

\paragraph{Instant Feedback}
An observation we made in \cref{sec:lectures}
extends to the practical part of the course:
feedback must come fast.
The benefit of prompt feedback is well supported in the literature~\cite{onlineengagement2,onlineengagement4}.
Again, an asynchronous Q\&A forum helps in this regard,
at least when dealing with questions of a general nature.
Problems specific to a student's submission (e.g.\ a bug or error in a proof),
however, must be fixed by the student herself as
\begin{enumerate*}[label=\arabic*)]
  \item it is a critical skill of computer scientists to discover bugs and
  \item code/proofs may not be shared before the submission deadline due to the grade bonus.
\end{enumerate*}

Automated tests can fill this gap:
they provide prompt feedback without giving away too much information (e.g.\ by only showing a failing input and expected output pair).
Needless to say, they are also crucial to
scale the homework system to a large number of students.
We describe our testing infrastructure in more detail in \cref{sec:tech_setup_test}.

However, we also let student assistants manually review all final submissions
in the first iteration of the course
to provide feedback not covered by automation,
in particular regarding code quality.
To our dismay, this feedback did very little and
most of it was probably ignored.
In part, this is because it took 1--2 weeks after each submission deadline to provide feedback to all students.
At that point, students had already moved on to a fresh set of exercises and were probably not motivated to revisit their old submissions.
Some may also only care about passing the tests
and are not particularly interested in feedback about code quality.

In our second iteration, we hence reallocated resources:
instead of grading submissions,
student assistants now supported us by creating engaging exercises
and offering new content (e.g.\ supervising workshops of industry partners)
while we focused on writing exhaustive tests with good feedback and extended our automated proof checking facilities (see \cref{sec:cyp}).
To provide at least some feedback on
code quality, we instructed students
to use a linter (see \cref{sec:tech_setup_test}).

We can report very positively on this decision:
we were able to offer a more diverse set of exercises and
had the resources to offer new content
while quality of code did not seem to suffer.
Indeed, the linter even seemed to increase students' awareness
to not only write correct code but also use good coding patterns.
This seems to be due to the fact that
\begin{enumerate*}[label=\arabic*)]
  \item the linter provides instant feedback and
  \item it visually highlights affected code fragments and provides quick fixes.
\end{enumerate*}

\paragraph{Competition and Awards}

Due to positive feedback,
we continued the tradition of running an opt-in weekly
programming competition as introduced in~\cite{next_1100}.
Each week, one homework assignment
was selected as a competition problem
and a criterion for ranking the submissions was fixed.
Participation was optional:
students could pass the exercise without optimising their code and submitting it to the competition.
The set of competition problems was diverse,
including code golf challenges,
optimisation problems,
game strategy competitions,
an ACM-ICPC-like programming contest,
and creative tasks like music composition
and computer art (see \cref{sec:selected_exercises}).
The top 30 entries received points
and were presented to the public on a blog\footnote{\url{https://www21.in.tum.de/teaching/fpv/WS20/wettbewerb.html} (WS20) and

\url{https://www21.in.tum.de/teaching/fpv/WS19/wettbewerb.html} (WS19)},
written using the ironic self-important third-person style established in previous semesters.

The overall top 30 students received awards at
a humorously organised award ceremony at the end of the semester.
We cooperated with industry partners
to offer prizes such as tickets to functional programming conferences,
Haskell workshops and programming books, as well as cash and material prizes.
This initial contact with industry partners
also sparked the idea to offer Haskell workshops
run by software engineers from industry in WS20 (explained further below).

The competition in WS20 greatly benefitted from incorporating the work of our student assistants:
At the beginning of the semester,
we brainstormed for competition ideas.
We then formed teams, each one being responsible for the
implementation of one idea to be published as a competition exercise during the semester.
This allowed us to create more extensive, diverse,
and practical exercises than in previous years,
where all tasks were created by the course organisers.

As reported in~\cite{next_1100},
we can confirm that the competition works extremely well to motivate talented students.
They go well beyond what is taught as part of the course when devising their competitive solutions.
Many of them became major drivers in
the team of student assistants in follow-up iterations.
Indeed, after offering the competition in WS19,
the number of applications for student assistant positions in WS20 more than doubled.
In each iteration of the course, 144 different students ranked among the top 30 of the week at least once.
We also received testimonies from students that even though they did not perform well (or participated at all) in the competition,
they nevertheless enjoyed the blog posts and advanced material discussed on it.

The competition combines multiple
effective engagement mechanisms~\cite{onlineengagement5,engagementproposals}:
it is challenging, often practical, humorous,
and integrates gamification aspects.
Despite the help of our student assistants,
running the competition remained enormously labour-intensive,
in particular the evaluation of submissions and
the writing of blog posts.
We envisage further help by student assistants in those regards,
cutting down the competition to a bi-weekly format,
or replacing it by more efficient mechanisms that motivate talented students.

\paragraph{Workshops with Industry Partners}
Many students at TUM have questioned the applicability and value of functional programming for real-world applications.
Obviously, there is not much use in us academics promising
them otherwise.
Instead, we had the idea to invite people from industry
to offer functional programming workshops about
practical topics not covered in our course.

In WS20,
we hosted three workshops on
\begin{enumerate*}[label=\arabic*)]
  \item design patterns for functional programs,
  \item networking and advanced I/O, and
  \item user interfaces and functional reactive programming.
\end{enumerate*}
We limited participation to 35 students for each workshop,
and to our delight, demand exceeded supply (more than 120 students applied).
Industry partners and workshop participants
alike
reported very positively to us.
In some cases,
workshops were even extended for multiple hours due to the great curiosity by students.
Moreover, organisational overhead was small:
we merely had to communicate the syllabus to our partners and coordinate time and place.
We envisage offering more workshops in future iterations and highly recommend this mechanism to other instructors.

\paragraph{Social Interactions}
Studies confirmed that the COVID-19 pandemic
worsened students' social life,
leading to higher levels of stress, anxiety, loneliness, and symptoms of depression~\cite{students_lockdown1}.
In WS20,
we hence investigated mechanisms
to foster social interaction and exchange between students
-- which also play an important pedagogical role in general~\cite{impact_social_interaction}.
Crucially,
there is no one-size-fits-all solution,
but multiple forms of social interaction
are needed to increase engagement~\cite{onlineengagement3,onlineengagement2}.

Firstly, we decided to employ pair-programming (groups of 3--4 students)
in our online tutorials.
The technical setup for this is described in \cref{sec:tech_setup_test}.
This not only made social interactions an integral part of the tutorial,
but also had positive effects on knowledge sharing.
We can report very positively on this policy:
it received 13 positive and 2 negative comments in the course evaluation form.

Secondly, we hosted two informal get-together sessions,
one at the beginning and one at the end of the semester.
Each session was joined by $\approx 50$ students.
We started with icebreaker sessions in breakout rooms,
randomly allocating students and at least one student assistant in each group.
We then opened thematic breakout rooms where students
could freely talk about a given topic.
Some preferred to talk about the course,
others had light-hearted conversations about university life,
yet others started to play online games.
All in all, we received very positive feedback for these sessions.

Thirdly, we organised an
ACM-ICPC-like programming contest
where students participated in teams,
followed again by a light-hearted get-together session
for participants.

\subsection{Technical Setup and Automated Assessment}\label{sec:tech_setup_test}

\paragraph{Automated Assessment}
In WS19, we used an improved version of
the testing infrastructure introduced in~\cite{next_1100}.
However, this system could neither manage online exams nor mark non-programming tasks.
We thus switched to a newly written open-source
tool developed at TUM called ArTEMiS~\cite{artemis}.
ArTEMis is a highly scalable, automated assessment management system and is programming language independent --
it only expects test runners to produce tests results
adhering to the Apache Ant JUnit XML schema.
It already offered support for a few imperative programming languages,
and we added support for Haskell.\footnote{It now also supports OCaml.}

As ArTEMis takes care of most things,
including automated test execution and score management,
and offers an exam mode and good support for grading non-programming tasks,
the only thing that was left to do was writing the testing code.
For the most part, we verified the results computed
by a student's submission by comparing them to those
computed by a sample solution written by us.
In some cases, we also tested for efficiency using timeouts.
Our tests were powered by the following libraries:
\begin{enumerate}
  \item QuickCheck \cite{quickcheck}:
  Although QuickCheck can automatically generate test data (using the typeclass \mintinline{Haskell}{Arbitrary}),
  most tests and benchmarks used custom input generators.
  This was necessary to increase coverage and eliminate non-applicable inputs for tests with preconditions.
  We also used custom shrinkers to provide better feedback to students in case of a failure.
  In both cases, the flexible combinators provided by QuickCheck made this a straightforward task.
  \item SmallCheck \cite{smallcheck}: The exhaustive testing facilities provided by SmallCheck mainly served
    as a complementary tool that provided small counterexamples for, in many cases, obvious deficiencies.
  \item Tasty\footnote{\url{https://hackage.haskell.org/package/tasty}}: We used Tasty to put QuickCheck, SmallCheck, unit tests, and the checking of ``Check Your Proof'' proofs (see \cref{sec:cyp}) into one common framework that is capable of generating results interpretable by ArTEMiS.
  We used the unit testing facilities of Tasty to complement our tests with corner cases.
  Integration of proof checking was pleasantly straightforward:
  one only has to provide a suitable instance for Tasty's \mintinline{Haskell}{IsTest} typeclass.\footnote{The code can be found in the repository in \href{https://github.com/kappelmann/engaging-large-scale-functional-programming/tree/main/resources/io_mocking/typeclass}{resources/cyp\_integration/test/hs/Test.hs}}
  Moreover, Tasty supports timeouts for individual test cases.
  solving the issue of truncated test reports mentioned in~\cite{next_1100}.
\end{enumerate}

\paragraph{Development Environment and Online Tutorials}
In previous iterations,
there were no recommendations for students regarding the development environment they should use for the practical part of the course.
However, due to the COVID-19 pandemic,
this was no longer an option:
we needed a way for students to share their code in read and write mode
to other peers (pair-programming) and teaching assistants (for feedback purposes) during online tutorials.
Moreover, as explained in \cref{sec:engagement},
we wanted students to use a linter.
Finally, we had negative experiences with students
installing no compiler at all
and instead (mis)using our submission server as a compiler backend.
We thus introduced a strict policy
for the technical setup to be used during
the tutorials.
Detailed installation instructions can be found online\footnote{\url{https://www21.in.tum.de/teaching/fpv/WS20/installation.html}}.
Here we briefly list the key components and our experiences:
\begin{enumerate}
  \item IDE: We used VSCodium\footnote{VSCodium provides free open-source software binaries of VSCode \url{https://vscodium.com/}} due to its cross-platform support,
    rich library of extensions,
    widespread adoption,
    and free open-source software philosophy.
    We did not receive any negative feedback by students
    and, besides few exceptions, no major installation problems were reported.
  \item Build and dependency manager: We used Stack\footnote{\url{https://www.haskellstack.org/}} for these purposes.
    Since Stack provides curated sets of packages and compiler versions that are checked for compatibility,
    deterministic builds are guaranteed.
    Students hence showed little struggle to compile and execute their programs.
  \item Linter: We used HLint\footnote{\url{https://github.com/ndmitchell/hlint}}, which, among other things,
    provides suggestions for alternative functions
    and simplified code and spots redundancies.
    Students reacted curiously and positively to these suggestions
    and discussed them vividly during pair-programming sessions.
  \item API search engine: Due to Haskell's strong type system,
    searching its API by type signature often returns better results than searching for function names.
    For this purpose, we introduced Hoogle\footnote{\url{https://hoogle.haskell.org/}} to our students and let them install an extension that integrates Hoogle searches into their IDE.
    Unfortunately, we have no data to report on whether this enriched students' programming~experiences.
  \item Real-time collaboration: We used VSLiveShare\footnote{\url{https://visualstudio.microsoft.com/services/live-share/}} to run our pair-programming sessions (groups of 3--4 students).
    We can report positively on its connection stability and usability.
    Unfortunately, the plugin requires a Microsoft or GitHub account,
    but since almost all students
    already signed up to the latter before,
    this requirement passed uncontroversially.
    Nevertheless, for future iterations,
    we plan to investigate open-source alternatives.
\end{enumerate}

All in all, we can report positively on the setup.
Students were more knowledgeable
about their programming environment (compilation, dependency management, etc.) than in previous iterations.
Misappropriation of the submission server as a compiler backend also stopped.

\subsection{Selected Exercises and Tools}\label{sec:selected_exercises}

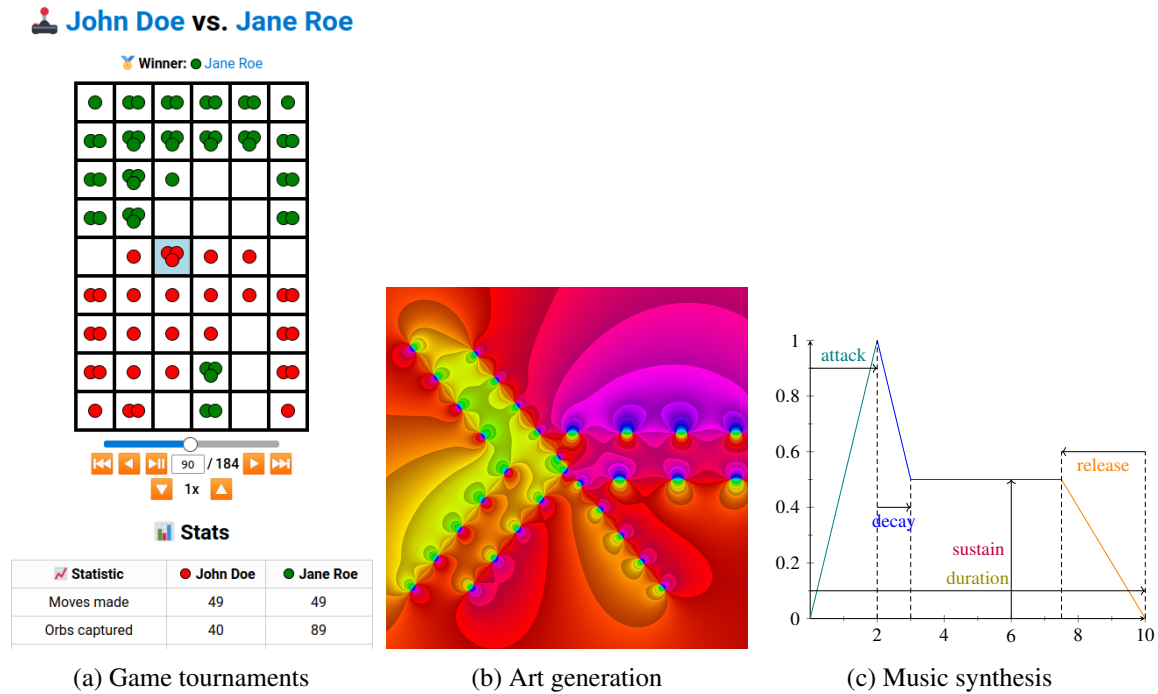
\begin{figure*}[t!]

\centering
\begin{subfigure}[t]{0.3\textwidth}
  \centering
  \includegraphics[width=\linewidth]{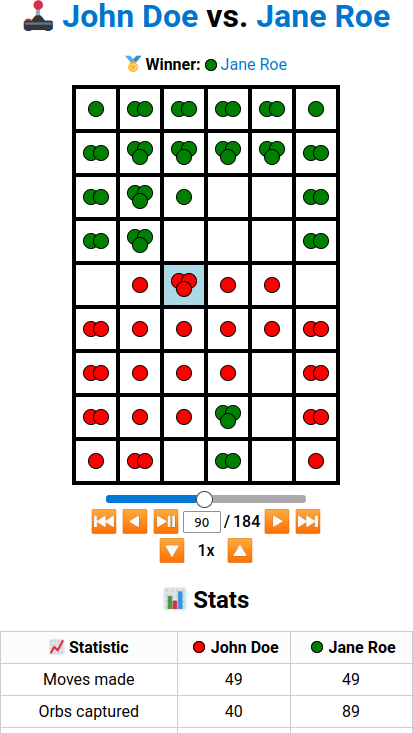}
  \caption{Game tournaments}
  \label{fig:chainreaction}
\end{subfigure}%
~
\begin{subfigure}[t]{0.3\textwidth}
  \centering
  \includegraphics[width=\linewidth]{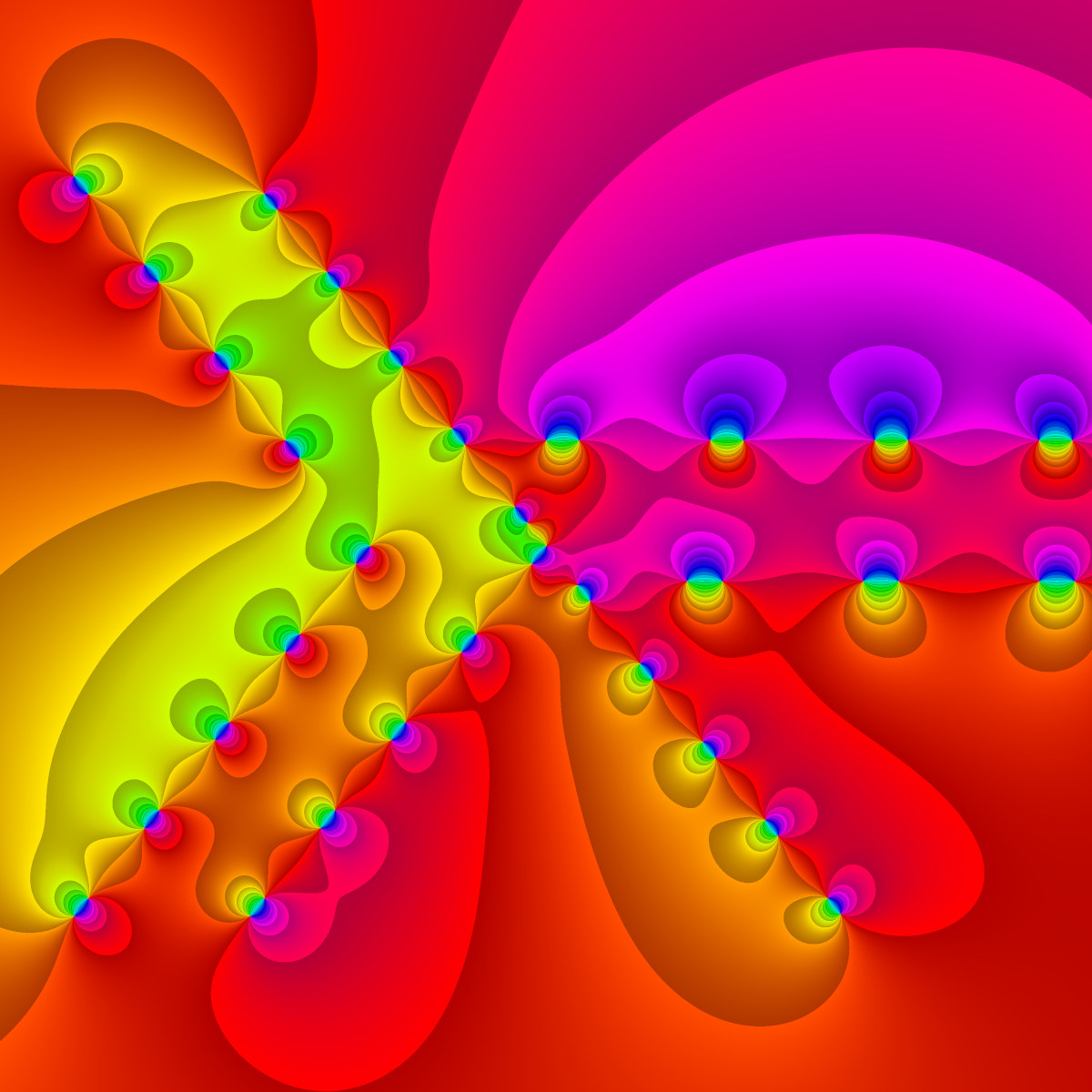}
  \caption{Art generation}
\end{subfigure}
~
\begin{subfigure}[t]{0.3\textwidth}
  \centering
  \begin{tikzpicture}[scale=0.65]
  \begin{axis}[
    minor tick num=1,
    samples=120,
    axis y line=left,
    axis x line=middle
    ]
    \addplot+[domain=0:10, mark=none, teal]
      coordinates {(0,0) (2, 1)};
    \addplot+[domain=0:10, mark=none, blue]
      coordinates {(2,1) (3, 0.5)};
    \addplot+[domain=0:10, mark=none, purple]
      coordinates {(3, 0.5) (7.5, 0.5)};
    \addplot+[domain=0:10, mark=none, orange]
      coordinates {(7.5, 0.5) (10, 0)};
    \addplot[densely dashed] coordinates {(2,0) (2,1)};
    \addplot[densely dashed] coordinates {(3,0) (3,0.5)};
    \addplot[densely dashed] coordinates {(7.5,0) (7.5,0.6)};
    \addplot[densely dashed] coordinates {(10,0) (10,0.6)};
    \addplot[->] coordinates {(0,0.9) (2,0.9)} node[midway, above, teal] {attack};
    \addplot[->] coordinates {(2,0.4) (3,0.4)} node[midway, below, blue] {decay};
    \addplot[->] coordinates {(6,0) (6,0.5)} node[midway,left, purple] {sustain};
    \addplot[<-] coordinates {(7.5,0.6) (10,0.6)} node[midway, below, orange] {release};
    \addplot[->] coordinates {(0,0.1) (10,0.1)} node[midway, above, olive] {duration};
  \end{axis}
  \end{tikzpicture}
  \caption{Music synthesis}
\end{subfigure}
\caption{Examples of exercises created as part of the course}
\end{figure*}

Many students at TUM have questioned the
applicability and usefulness
of functional languages after completing
the mandatory functional programming course.
We believe this is mainly due to two reasons:
\begin{enumerate*}[label=\arabic*)]
  \item introductory courses often stick
to simple algorithmic or mathematically inspired challenges and
\item side-effects (in particular I/O)
are often introduced rather late
in functional programming courses.
\end{enumerate*}

While we were able to introduce I/O midway through the course,
introducing it even earlier appeared difficult to us:
we think that students would be confused
if a ``special'' \mintinline{Haskell}{IO} type and \mintinline{Haskell}{do} notation were to
be introduced before they are comfortable
with the basic features of functional languages.
We thus focused on the other issue
and created exercises that go beyond
simple terminal applications.
Designing and implementing such exercises,
however, is labour-intensive.
As mentioned in \cref{sec:engagement},
we thus decided to reallocate resources and
let our student assistants help us with this work
rather than providing feedback for homework submissions.

This turned out to be a very fruitful idea:
the quality of our student assistants' work was often way above what we expected.
The one difficulty we faced was the mediocre quality of
tests written by most assistants
since they only had the rudimentary knowledge of QuickCheck taught as part of the course.
We thus hosted a workshop for them that explained
our testing infrastructure and best-practice
patterns when writing tests.
The quality of tests significantly increased following this workshop,
though we still had to polish them before publication.

We next introduce a few exercises and tools
that were created as part of the course.
They are available in this article's repository\footnote{\url{https://github.com/kappelmann/engaging-large-scale-functional-programming}},
next to our other exercises, including
a music synthesiser framework,
a turtle graphics framework,
an UNO framework,
and guided exercises for DPLL and resolution provers.

\paragraph{Game Tournament Framework}
It has become a course tradition to run a game tournament over the Christmas break.
In this tournament, students are tasked with writing an AI for a board game that competes against the AIs of their fellow students.
To pass the homework sheet, it suffices to implement a basic strategy, but to score well in the competition, students came up with quite sophisticated strategies in past years.
The framework allows students to use statefulness and randomisation, so that there are few limits to the students' creativity.

The tournament runs continuously for 2--3 weeks and the results are displayed on a website (see \cref{fig:chainreaction} for an example from WS20).
Students thus get reasonably quick feedback on how their strategy performs,
which keeps them engaged and allows them to improve their strategy iteratively.
The tournament became a popular feature of the course with 182 participating students in WS19 and 220 in WS20.

In our repository, we provide the framework along with code specific to the game from WS20, which is based on Chain Reaction\footnote{\url{ https://brilliant.org/wiki/chain-reaction-game/}}.
It runs a round-robin tournament,
collecting statistics for each game and player.
Instructions for adapting the framework to a different game can also be found in the repository.

\paragraph{Programming Contest Framework}\label{sec:contest}
To foster social interaction and diversify the bonus system,
we hosted an ACM-ICPC-like programming contest.
In such contests, students
participate in teams of 2--3,
solving as many programming challenges as possible in a given time frame,
and can check their ranking on a live scoreboard.

We found existing solutions
to run such contests too complex for our purpose
and hence created a lightweight alternative.
Our framework continuously receives test results,
computes each team's score,
and displays the live scoreboard and task instructions on a website.
It is agnostic to the programming language and test runner used.
It expects tests results adhering to the Apache Ant JUnit XML schema,
but modifying it to support other formats is straightforward.
Deployment instructions can be found in this article's repository.

We ran an online iteration of the contest in WS20,
again using ArTEMiS as a test runner.
Teams were cooperating on their platform of choice
and were able to ask for clarifications on a dedicated online channel.
Our experiences are positive:
27 teams participated in the contest
and most stayed for the social hangout following it.
Using our framework,
the technical setup of the contest requires little time.
Some significant time, however,
must be spent on setting up the challenges,
tests, and solutions,
though plenty of them may be found
online by searching for other contests,
which one then may modify and reuse.
We recommend offering such contests
to programming course instructors in general.

\paragraph{I/O-Mocking Library}
As discussed in \cref{sec:tech_setup_test}, we primarily use QuickCheck to automatically assess homework submissions.
This raises the question how monadic I/O in Haskell can be tested.
Since we do not want to actually execute the side effects that the submitted code produces,
the obvious solution is to use a mocked version of Haskell's \mintinline{Haskell}{IO} type.

A standard approach to mock \mintinline{Haskell}{IO}, which is put forward by packages such as \texttt{monad-mock}\footnote{\url{https://hackage.haskell.org/package/monad-mock}} and \texttt{HMock}\footnote{\url{https://hackage.haskell.org/package/HMock}}, is to first extract the side effects that are required for a certain computation into a new typeclass.
Since a typeclass allows multiple instantiations,
we can then provide one instantiation that actually executes the side effects on the machine
and another one that just modifies a mocked version of the environment.
For example, to implement a function that copies a file,
we need two operations:
one for reading a file and one for writing a file.
\begin{minted}{Haskell}
import qualified Prelude
import Prelude hiding (readFile, writeFile)

class Monad m => MonadFileSystem m where
  readFile :: FilePath -> m String
  writeFile :: FilePath -> String -> m ()
\end{minted}
The implementation is straightforward.
\begin{minted}{Haskell}
copyFile :: MonadFileSystem m => FilePath -> FilePath -> m ()
copyFile source target = do
  content <- readFile source
  writeFile target content
\end{minted}

Due to the definition of \mintinline{Haskell}{MonadFileSystem}, the instance for \mintinline{Haskell}{IO} is trivial.
The mocked version can be implemented as a map from file names to file contents wrapped by the \mintinline{Haskell}{State} monad transformer to make it mutable.
We omit this instantiation of \mintinline{Haskell}{MonadFileSystem} for brevity.
Testing \mintinline{Haskell}{copyFile} is now as simple as checking whether the state of the file system is as expected after executing the function.
An example that includes the instance \mintinline{Haskell}{MonadFileSystem (State MockFileSystem)} and a test can be found in the repository in \href{https://github.com/kappelmann/engaging-large-scale-functional-programming/tree/main/resources/io_mocking/typeclass}{resources/io\_mocking/typeclass}.
\begin{minted}{Haskell}
instance MonadFileSystem IO where
  readFile = Prelude.readFile
  writeFile = Prelude.writeFile

data MockFileSystem = MockFileSystem (Map FilePath String)
instance MonadFileSystem (State MockFileSystem) where
  readFile = ...
  writeFile = ...
\end{minted}

While this approach is sufficient for many use cases,
it lacks one important property: transparency.
More specifically, the code submitted by students must contain or import \mintinline{Haskell}{MonadFileSystem} and the signatures of terms that use \mintinline{Haskell}{IO} must be adapted.
This is especially problematic because the lecture introduces \mintinline{Haskell}{IO} without mentioning monads.
Other approaches to test I/O
face similar issues~\cite{iotest1,iotest2}.

Instead of modifying existing code,
we delay the mocking to the compilation stage.
We achieve this with a mixin that replaces the \mintinline{Haskell}{IO} module of a submission with a mocked version.
The mocked \mintinline{Haskell}{IO} type can be realised similarly to above mock file system.
However, to achieve full transparency,
we not only need a file system but also handles, such as standard input and output, as well as a working directory.

All these aspects of the machine state are summarised in the type \mintinline{Haskell}{RealWorld} as seen below.
Crucially, the type also contains a mock user,
represented by a computation of type \mintinline{Haskell}{IO ()},
which interacts with a student's submission;
that is, the user generates the input for and reads the output of the student's submission.
For brevity, we do not show the full type here.
\begin{minted}{Haskell}
data RealWorld = RealWorld {
  workDir :: FilePath,
  files :: Map File Text,
  handles :: Map Handle HandleData,
  user :: IO (),
  ...
}
\end{minted}
Again, this type is wrapped by the \mintinline{Haskell}{State} monad transformer as well as two additional transformers \mintinline{Haskell}{PauseT} and \mintinline{Haskell}{ExceptT} in order to form the mocked \mintinline{Haskell}{IO} type.
\begin{minted}{Haskell}
newtype IO a =
  IO { unwrapIO :: ExceptT IOException (PauseT (State RealWorld)) a }
\end{minted}

While the transformer \mintinline{Haskell}{ExceptT} simply adds I/O exceptions,
such as errors for insufficient permissions,
the purpose of \mintinline{Haskell}{PauseT} is not obvious.
To understand its role, consider the following simple program that reads the user's name and greets them.
\begin{minted}{Haskell}
module Hello where

main = do
 name <- getLine
 putStrLn $ "Hello " ++ name
\end{minted}
In a normal (non-mocked) execution of the program, the program blocks and waits for input when \texttt{getLine} is called.
If our mocked \mintinline{Haskell}{IO} type would only consist of a state monad,
all the input to the program would have to be passed in one monolithic unit.
However, programs may consume input multiple times.
We thus need a way to suspend the program every time a blocking operation is called and transfer control over to our mock user.
The mock user then reacts to the output of the program and generates the input that the program is waiting for.
When the user is done, it yields and the program of the student is resumed.

These considerations lead us to the monad below,
consisting of two operations:
The first one pauses execution and the second one runs a computation of the monad until either \texttt{pause} is called or the computation finishes.
In the former case,
\texttt{stepPauseT} returns a \mintinline{Haskell}{Left c} where \texttt{c} represents the rest of the computation, i.e.\ the part of the computation that is executed when resuming.
Otherwise, the final result \texttt{r} of the computation is returned as \mintinline{Haskell}{Right r}.
It should be noted that the pause monad is an instance of the more general coroutine monad as provided by the \texttt{monad-coroutine}\footnote{\url{https://hackage.haskell.org/package/monad-coroutine}} package.
For the implementation details of the corresponding monad transformer \mintinline{Haskell}{PauseT}, we refer to the repository.
\begin{minted}{Haskell}
class Monad m => MonadPause m where
  pause :: m ()
  stepPauseT :: m a -> m (Either (m a) a)
\end{minted}

We exemplify the mechanics of the mocking framework with a simple test of above \mintinline{Haskell}{main} function.
To this end, we first implement a mock user that takes a name and supplies it to the standard input of \mintinline{Haskell}{main}.
The user then reads the output of the program and checks whether it printed the expected greeting.
In the QuickCheck property \mintinline{Haskell}{prop_hello},
we evaluate the interaction between the mock user and the program with \mintinline{Haskell}{Mock.evalIO} on \mintinline{Haskell}{Mock.emptyWorld}, a minimal \mintinline{Haskell}{RealWorld} that contains no files and only the absolutely necessary handles: standard input, standard output, and standard error.
The interaction itself sets the user to \texttt{user s},
then executes the \mintinline{Haskell}{main} function, and finally runs the user to completion.
\begin{minted}{Haskell}
import qualified Mock.System.IO.Internal as Mock
import qualified Hello as Sub

user :: String -> Mock.IO ()
user s = do
  Mock.hPutStrLn Mock.stdin s
  output <- Mock.hGetLine Mock.stdout
  when (output /= ("Hello " ++ s))
    (fail $ "\nExpected:\n" ++ "Hello " ++ s
      ++ "\nActual:\n" ++ output ++ "\n")

prop_hello = forAll (elements ["Karl", "Friedrich", "Rosa"]) $ \s ->
  Mock.evalIO (Mock.setUser (user s) >> Sub.main >> Mock.runUser)
              Mock.emptyWorld
\end{minted}
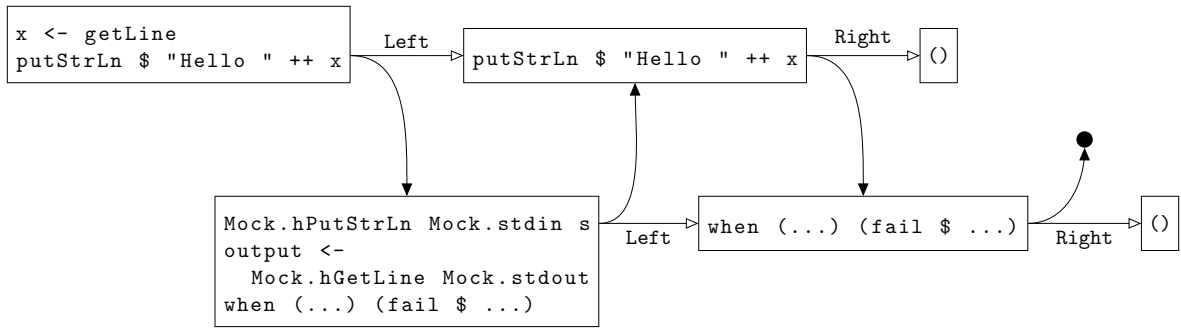
\begin{figure}[t!]
  \centering
\scalebox{0.75}{
\begin{tikzpicture}[cnode/.style={draw,rectangle,minimum height=2.5em}, >=stealth]
  \node[cnode] (P1) {
\begin{lstlisting}[style=Haskell]
x <- getLine
putStrLn $ "Hello " ++ x
\end{lstlisting}};
  \node[cnode, right=2cm of P1.south east, anchor=south west] (P2) {
\begin{lstlisting}[style=Haskell,firstline=2]
putStrLn $ "Hello " ++ x
\end{lstlisting}};
  \node[cnode, right=2cm of P2.south east, anchor=south west] (P3) {\texttt{()}};

  \path (P1.south east) -- coordinate (MP12) (P2.south west);
  \path (P2.south east) -- coordinate (MP23) (P3.south west);
  \node[cnode, below=2cm of MP12, anchor=north] (U1) {
\begin{lstlisting}[style=Haskell]
Mock.hPutStrLn Mock.stdin s
output <-
  Mock.hGetLine Mock.stdout
when (...) (fail $ ...)
\end{lstlisting}};
  \node[cnode, below=2cm of MP23, anchor=north] (U2) {
\begin{lstlisting}[style=Haskell,firstline=4]
when (...) (fail $ ...)
\end{lstlisting}};
  \node[cnode, right=2cm of U2.north east, anchor=north west] (U3) {\texttt{()}};

  \path (U2.east |- U3) -- coordinate (MU23) (U3);
  \path (P3 -| MU23) -- node[shape=circle, fill=black, scale=0.8] (END) {} (MU23);

  \begin{scope}[>=open triangle 45]
    \draw[->] (P1.east |- P2) -- node[above] {\texttt{Left}} (P2);
    \draw[->] (P2.east |- P3) -- node[above] {\texttt{Right}} (P3);
    \draw[->] (U1.east |- U2) -- node[below] {\texttt{Left}} (U2);
    \draw[->] (U2.east |- U3) -- node[below] {\texttt{Right}} (U3);
  \end{scope}

  \begin{scope}[>=triangle 45]
    \draw[->] (P1.east |- P2) to[out=0, in=90] (U1);
    \draw[->] (P2.east |- P3) to[out=0, in=90] (U2);
    \draw[->] (U1.east |- U2) to[out=0, in=-90, looseness=0.85] (P2);
    \draw[->] (U2.east |- U3) to[out=0, in=-90] (END);
  \end{scope}

\end{tikzpicture}
}
\caption{
  Interaction between the mock user and the student's submission.
  White arrows indicate the return value of \texttt{stepPauseT} and black arrows indicate transfer of control.
  The black dot signifies the end of the interaction.\label{fig:iomocking}
}
\end{figure}

\cref{fig:iomocking} illustrates the evaluation steps of \mintinline{Haskell}{Mock.evalIO}.
Note that there are two blocking operations (i.e.\ operations that call \texttt{pause} internally),
namely \texttt{getLine} and \mintinline{Haskell}{Mock.hGetLine Mock.stdout}.
When \mintinline{Haskell}{Mock.evalIO} encounters any such operation,
it transfers control between the user and the program as indicated by the black arrows.
Control is also transferred if the computation runs until completion without meeting a \texttt{pause}.
The horizontal axis with white arrows illustrates the return values of \texttt{stepPauseT}.
Focusing on \mintinline{Haskell}{main},
we see that \texttt{stepPauseT}
returns the remaining computation \mintinline{Haskell}{putStrLn $ "Hello " ++ x} wrapped in a \mintinline{Haskell}{Left}
when encountering \texttt{getLine}.
After the user provides the input for \texttt{getLine} and yields, the \mintinline{Haskell}{main} function prints the greeting and finishes with the result \mintinline{Haskell}{Right ()}.
Similarly, the user is blocked on \mintinline{Haskell}{Mock.hGetLine},
which means that the remaining computation only consists of the \texttt{when} check, which is executed as soon as \mintinline{Haskell}{main} is done.
This explains why we need to run \mintinline{Haskell}{Mock.runUser} after \mintinline{Haskell}{Sub.main} since the crucial \texttt{when} check would never be executed otherwise.

All in all, the mocking framework lets us uniformly test student submissions with common frameworks like QuickCheck and SmallCheck regardless of whether they contain I/O effects.

\section{Check Your Proof by Example}\label{sec:cyp}
Besides functional programming,
the course also dealt with the verification of functional programs.
Even though we only consider a strict and total subset
of Haskell for our proofs,
equational reasoning together with induction
(and case analysis)
is already sufficient to prove interesting properties.
Since ``fast and loose reasoning is morally correct''~\cite{fast_and_loose},
valid properties in this sub-language carry over to Haskell.
Simple inductive proofs lend themselves well to be automatically checked and, as announced in~\cite{next_1100}, a tool called ``Check Your Proof'' (CYP for short) was developed at our lab by Durner and Noschinski\footnote{\url{https://github.com/noschinl/cyp}}.

The first example of such a proof presented in the lecture is the proof of associativity of the append-operator for lists.
Proving this example in CYP first requires us
to define the data type of lists.
\begin{lstlisting}[style=cyp]
  data List a = [] | a : List a
\end{lstlisting}
Now we can define the infix append-operator \lstinline[style=cyp]!++!
\begin{lstlisting}[style=cyp]
  [] ++ ys = ys
  (x : xs) ++ ys = x : (xs ++ ys)
\end{lstlisting}
and state the goal
\begin{lstlisting}[style=cyp]
  goal xs ++ (ys ++ zs) .=. (xs ++ ys) ++ zs
\end{lstlisting}
where all variables are implicitly universally quantified.
The listings above describe the background theory of the proof.
The theory is fixed in a file and provided
to the students.
The students then supply the proof in a separate file.
In our case, we proceed to prove the statement by structural induction on \lstinline[style=cyp]!xs!.
\begin{lstlisting}[style=cyp]
  Lemma: xs ++ (ys ++ zs) .=. (xs ++ ys) ++ zs
  Proof by induction on List xs
    Case []
      To show: [] ++ (ys ++ zs) .=. ([] ++ ys) ++ zs
      Proof
                        [] ++ (ys ++ zs)
        (by def ++) .=. ys ++ zs
        (by def ++) .=. ([] ++ ys) ++ zs
      QED
\end{lstlisting}

In above listing, the first \lstinline[style=cyp]!Proof! marks the beginning of an inductive proof whereas the second \lstinline[style=cyp]!Proof!, which has no further arguments, starts an equational proof.
While CYP allows to arbitrarily nest proofs
by case analysis or induction,
the innermost goal must always be discharged by an equational proof.
An equational proof is a chain of equations that rewrite the left-hand side of the current goal to the right-hand side.
The user has to justify each step by a corresponding equation that yields the right-hand side
when applied to the term or a subterm on the left-hand side.
Note that in the example, \lstinline[style=cyp]!(by def ++)! refers to either one of the defining equations of \lstinline[style=cyp]!++!
and CYP will check if any of them justifies the current step.

Now, consider the inductive case.
\begin{lstlisting}[style=cyp]
    Case x : xs
      To show: (x : xs) ++ (ys ++ zs) .=. ((x : xs) ++ ys) ++ zs
      IH: xs ++ (ys ++ zs) .=. (xs ++ ys) ++ zs
      Proof
                          (x : xs) ++ (ys ++ zs)
        (by def ++)   .=. x : (xs ++ (ys ++ zs))
        (by IH)       .=. x : ((xs ++ ys) ++ zs)

                          ((x : xs) ++ ys) ++ zs
        (by def ++)   .=. (x : (xs ++ ys)) ++ zs
        (by def ++)   .=. x : ((xs ++ ys) ++ zs)
      QED
  QED
\end{lstlisting}
Again, CYP requires the user to be explicit: the goal in each inductive case and the induction hypotheses have to be spelled out.
Note that CYP also allows one to start rewriting from the left-hand side as well as the right-hand side of the goal.

The lecture goes beyond proofs by structural induction and introduces proofs by extensionality, case analysis, and computation induction,
all of which CYP supports with some conditions applying as we will describe shortly.
Computation induction was not supported in the original version by Durner and Noschinski but was only introduced in a fork by us in WS20\footnote{\url{https://github.com/lukasstevens/cyp}}.
CYP also allows proving named auxiliary lemmas,
which is useful to modularise proofs
and often necessary to prove generalisations of a given goal.

The simplicity of CYP both in its usage and its implementation comes with some caveats:
\begin{itemize}
    \item The version of CYP used by us is untyped and thus unsound if the background theory contains multiple types as demonstrated in~\cite{cyp_holes}: given a singleton type \lstinline[style=cyp]!data U = U! in the background theory,
      one can prove \lstinline[style=cyp]!x .=. y! by case analysis.
        In an untyped environment, one can then use this lemma to prove equality between any two terms.
    \item Barring soundness issues due to a lack of types, one also needs to ensure that all function definitions are total and that their patterns do not overlap.
    \item Computation induction is unsound if there are recursive calls in branches of an if-then-else expression.
      This is because the induction hypotheses would have to be conditional rewrite rules in such cases, which CYP does not support.
\end{itemize}

Renz et al.~\cite{cyp_holes} discuss the inner workings of CYP in detail and develop an extension that introduces types,
thus solving the first issue.
They also made it possible to leave holes in proofs and in expressions, which then have to be filled in by students.
Solving the other issues, however, would incur additional effort.
We think that CYP should be put on a stronger foundation,
such as higher-order logic (HOL),
without compromising its simplicity from a user perspective.
One possibility would be to rewrite CYP as a frontend to Isabelle/HOL~\cite{isabelle}.
Since HOL is a typed logic, it would solve the first issue.
The latter issues could be resolved using Isabelle's function package~\cite{isabelle_functions}.
A stronger foundation would also allow one to develop new extensions for CYP more confidently.

Our teaching experience with CYP has been very positive.
CYP proof checking is quick and can easily be integrated in the testing framework (Tasty) that we used for our programming exercises.
This allowed us to deal with programming and proof exercises in a uniform and scalable way.
CYP was also generally liked by our students:
In WS19,
we received two positive comments and one negative comment without asking for feedback on CYP.
When asked explicitly about their thoughts on CYP in WS20,
the students answered with 18 positive comments and three negative comments.
The students liked the instantaneous feedback that CYP provides,
which helped them to deepen their understanding of inductive proofs at their own pace.
The main criticism of CYP was the lack of documentation.
A CYP-cheatsheet was later added to the repository of CYP to improve the situation.

In our exams, two exercises (out of $\approx 8$) were concerned with inductive proofs.
The first exercise was a straightforward proof by structural induction while the second one required more effort, e.g.\ a generalisation of the goal or a slightly less trivial computation induction.
We did not require the students to stick to CYP's syntax in the exam, but we urged them to follow a similar structure,
which worked very well overall and simplified the grading.
However, we have no data on whether CYP also improves students' understanding of induction.

\section{Exams}\label{sec:exam}
In previous iterations of the course, exams were paper-based by necessity as university regulations made digital exams unfeasible.
Due to the COVID-19 pandemic, however, remote exams
were allowed for the repeat exam in WS19 and both exams in WS20.
In the following, we outline how we adapted our exam process to this new reality and the advantages of our approach for students and staff.

\paragraph{Workflow for Students}

We chose the ArTEMiS~\cite{artemis} platform, previously described in \cref{sec:tech_setup_test}, for the exams.
In WS19, students were mostly unfamiliar with ArTEMiS,
but in WS20, they had been submitting their homework using the same workflow as during the exam.
Students were able to check out individual exam questions from a git repository and then work on them using their preferred programming environment.
In addition to programming exercises, the exams included theory questions and proofs, which were submitted as text using an online editor.

The exams were open-book, including online resources,
but posting questions to chats, forums, etc.\ was prohibited.
Usage of third-party code had to be cited using comments.
We think that the combination of a programming environment customised to the individual student's taste and access to online resources comes close to how programming is done in practice.
In contrast, previous exams were purely paper-based.
This limited the scope of exercises we were able to pose
since both writing and grading programs on paper is highly labour-intensive.
Moreover, we were able to expect a higher standard of correctness
since minor errors, in particular in syntax, are excusable when programming on paper but less so when students can test, or at least run, their programs before submission.

\paragraph{Grading}
Grading hundreds of exams on paper is a huge undertaking.
In previous iterations,
it took 10--20 people about 4--5 days to grade an exam.
Additionally, grading programming exercises on paper is error-prone, and it is unfeasible to digitise every paper submission and run it through a compiler.

Using ArTEMiS, we were able to use automated testing for programming exercises.
In most cases, we marked submissions passing all tests as fully correct.
Manual grading was only needed for submissions that failed some tests or did not compile.
In our view, manual grading is still necessary in those cases because it is generally not possible to write tests that are fine-grained to such a degree that every potential source of error is recognised and scored proportionally.
But also manual grading was simplified since test results provide valuable guidance when checking for errors in code.
Moreover, one could re-run tests after fixing compilation and semantic errors and deduct points accordingly.
As a result, we were able to grade an exam in 3 days.

\paragraph{Online Review}
Once graded, students have the right to review their
exams to check for errors in the grading process.
For paper-based exams at TUM, students usually had to make an
appointment for a time-slot to review their exam under
the supervision of teaching staff.
In large courses, this imposed a significant organisational overhead for both staff and students.

Using ArTEMiS, every student could simply review their exam and submit complaints through an online interface.
Following the review period,
the complaints were approved or rejected by the instructors.
This resulted in a vastly increased proportion of students reviewing their exam,
thus ensuring a higher degree of fairness in the grading process.
In addition, the quality of feedback for students improved
since
they now also got to see the test results of their submission next to the in-line grading comments made by
teaching staff.

\paragraph{Cheating}

Regulations at TUM allowed both supervised or unsupervised remote exams.
We considered supervision but ultimately decided against it. Were we to supervise the exam ourselves,
each staff member would have to keep track of 20--30 students via their webcam in a video conferencing software.
We feel that this would hardly ensure against cheating
since, for example, students would still be able to take advice from someone out of view of the camera or communicate with others online (it does not seem feasible to simultaneously supervise the camera and screen-share of 20--30 students).
There are also commercial options for exam supervision,
but it is unclear how effective they are at preventing cheating, and the options we reviewed raised significant privacy concerns.

We thus decided to rely on an honour system for our exams.
Beyond the honour pledge,
we also checked for plagiarism with Moss.
This turned up only three cases of conclusive plagiarism.
We suspect and accept that there were likely more cases of cheating that we could not catch,
so we aim to improve our anti-cheating measures in future iterations.
We also created 3--4 slight variations of each exam question
in order to make plagiarism slightly more onerous. ArTEMiS supports the creation of these variants and assigns them randomly during the exam.

\section{Related Work}\label{sec:related_work}

The literature on student engagement in higher-education is vast,
and we are not in the position to give a thorough account of it.
Some general accounts of and suggestions for
student engagement can be found in~\cite{student_engagement,engagementproposals}
while work specifically focusing on online courses can be found in~\cite{onlineengagement3,onlineengagement2,onlineengagement4,onlineengagement5,onlineengagement1}.
Many of the mechanisms described there were employed in our course.
As such, our work can be seen as a case study, testing the hypotheses framed in these works.

Much work has been spent on the
automated assessment of programming exercises.
A recent overview can be found in~\cite{automatedassessment}.
We do not to provide another such software package but
describe how various existing solutions
can be put together to
address the challenges laid out in~\cref{sec:intro}.

To the best of our knowledge,
there are only few papers about converting physical computer science classes to virtual ones
and none that deal with all the challenges outlined by us.
Two experience reports using imperative languages can be found in~\cite{largeprogrammingclass,onlinecourse1}.
While they use a project-based learning approach,
that is students work in teams on graded projects,
our students had to submit their homework individually.
We fostered social interaction and team-based learning
in the context of the tutorials, workshops,
and programming contests instead.
We believe that this strikes a good balance
between individual work and team work.

There are several studies and experience reports that corroborate some of our findings, for example
\begin{itemize}
  \item increasing engagement using live programming~\cite{fp_first_year_risks_benefits,livecoding1,livecoding2}, pair-programming~\cite{engagingprogramming,teaching_fp_first_year}, and gamification~\cite{soccerfun,teaching_fp_glossy_games,teaching_fp_macedonian},
  \item letting students ask questions during the lecture via chat~\cite{increase_interest_fp},
  \item using automated grading for prompt feedback~\cite{teaching_fp_glossy_games,teaching_art_fp_automated,teaching_fp_macedonian}, and
  \item introducing I/O early (without explaining monads)~\cite{fp_first_year_risks_benefits,haskell_school_hudak,teaching_fp_chalmers} and establishing connections to industry by organising events~\cite{teaching_fp_chalmers} to demonstrate ``real-world'' applicability.
\end{itemize}

Various articles have been published about
specific tools to teach functional programming,
such as I/O testing~\cite{iotest2}
and visualisation of program evaluation~\cite{steppingocaml}.
While our work contributes to this realm,
it also describes more general strategies
to run engaging (functional) programming courses.

Finally,~\cite{next_1100}
introduces a previous iteration of our course,
focusing on the lecture material,
specific exercises, and a custom-made testing infrastructure.
Our objective, instead, is to provide strategies and resources to address
the challenges laid out in~\cref{sec:intro},
i.e.\ the drastic increase in student enrolments,
the challenges of online teaching,
and the difficulty of demonstrating the practicality of functional programming.

\section{Conclusion}\label{sec:conclusion}

As computer science departments continue to grow,
COVID-19 continues to spread,
and imperative programming appears to be the industry norm,
we functional programming educators
have to find ways to make our courses more scalable and
engaging, while demonstrating the elegance and usefulness of functional programming and having to adapt to a mix of teaching in physical and virtual space.
We hope our efforts not only inspired our 2000 students
but also other educators to take on these challenges.
The insights and resources presented in this article proved valuable to us and will hopefully also do to others.

For future iterations,
we plan to keep many things that we introduced during the pandemic and double down on some new engagement mechanisms such as the supplementary workshops.
One thing we want to improve is the
time we have to spend on the weekly competition,
for example by handing some of the workload to student assistants and creating fewer but more engaging exercises.
Another possibility we want to explore is to use more
competition exercises that can be automatically graded
and their results continuously be published on a website (like our Chain Reaction competition in \cref{sec:selected_exercises}).
This would also increase student engagement due to instant feedback.
We also plan to stick to the digital open-book exam format,
though as of yet, we have no good solution to prevent cheating.

\paragraph{Acknowledgements}
We want to thank all people involved in the course,
in particular Tobias Nipkow,
Manuel Eberl,
our student assistants,
the ArTEMiS development team,
our industry partners
Active Group,
QAware,
TNG Technology Consulting,
and Well-Typed,
and our 2000 Haskell students.
Finally, we thank the participants of TFPIE 2022
and the anonymous reviewers for their valuable feedback.

\bibliographystyle{eptcs}
\bibliography{sources}
\end{document}